\documentclass{article}
\usepackage{spconf,amsmath,graphicx,hyperref}
\usepackage[style=numeric, sorting=none]{biblatex}
\usepackage{enumitem}
\usepackage{caption}
\usepackage{amsmath,amssymb,amsfonts}
\usepackage{tipa}
\usepackage{booktabs}
\title{SSVD-O: Parameter-Efficient Fine-Tuning with Structured SVD for Speech Recognition}
\name{Pu Wang~$^{1}$ \qquad Shinji Watanabe~$^{2}$ \qquad Hugo Van hamme~$^{3}$}
  
  \address{$^{1, 3}$~KU Leuven, Department of Electrical Engineering, Leuven, Belgium \\
      $^{2}$~Carnegie Mellon University, Language Technologies Institute, Pittsburgh, PA, USA\\
      $^{1}$~pu.wang@esat.kuleuven.be, $^{2}$~shinjiw@ieee.org, $^{3}$~hugo.vanhamme@esat.kuleuven.be}
\begin{document}
\ninept
\maketitle
\begin{abstract}
Parameter-efficient fine-tuning (PEFT) is a scalable approach for adapting large speech foundation models to new domains. While methods such as LoRA and its state-of-the-art variants reduce adaptation costs, they typically allocate parameters uniformly across model subspaces, which limits their efficiency and scalability in speech applications. Building on our prior work, this paper introduces SSVD-Outer (SSVD-O), an extension of the structured SVD-guided (SSVD) fine-tuning method. SSVD-O combines input acoustic feature space–associated inner transformations with output semantic feature space–associated outer transformations to enable scalable and balanced adaptation. We conduct the first systematic analysis of parameter budget allocation across model subspaces in PEFT for automatic speech recognition (ASR), and investigate the trade-off between learning and forgetting under constrained resources. SSVD-O is benchmarked against LoRA, DoRA, PiSSA, and SSVD on domain-shifted ASR tasks, including child speech and regional accents, across model scales from 0.1B to 2B within the ESPnet framework. Experimental results show that SSVD-O consistently narrows the performance gap to full fine-tuning while improving generalization and mitigating catastrophic forgetting.
\end{abstract}
\section{Introduction}
\label{sec:intro}
Speech Foundation Models (SFMs), such as OpenAI’s Whisper~\cite{radford2023robust}, Google’s USM~\cite{zhang2023google}, the open-source Whisper-style OWSM~\cite{peng2023reproducing, peng2024owsm}, and NVIDIA’s Canary~\cite{puvvada2024less}, have demonstrated strong generalization across multilingual speech tasks following scaling laws~\cite{chen2025owls}. However, since SFMs are typically trained on generic, easily accessible corpora dominated by adult, healthy speech, standard accents, and mainstream languages, directly adapting them to low-resource but domain-shifted ASR tasks (e.g., child speech, regional dialects, or disordered speech) remains challenging~\cite{wang2023benefits, ying2025benchmarking}. Fine-tuning SFMs for downstream speech conditions has therefore become the dominant paradigm for automatic speech recognition (ASR). In such scenarios, mismatches in acoustic features, linguistic structures, and articulatory behaviors can significantly degrade performance, while full fine-tuning becomes computationally expensive for billion-scale models. How to efficiently adapt large-scale models with limited in-domain data remains an open question.
\begin{figure}[!htbp]
\centerline{\includegraphics[width=1.0\linewidth]{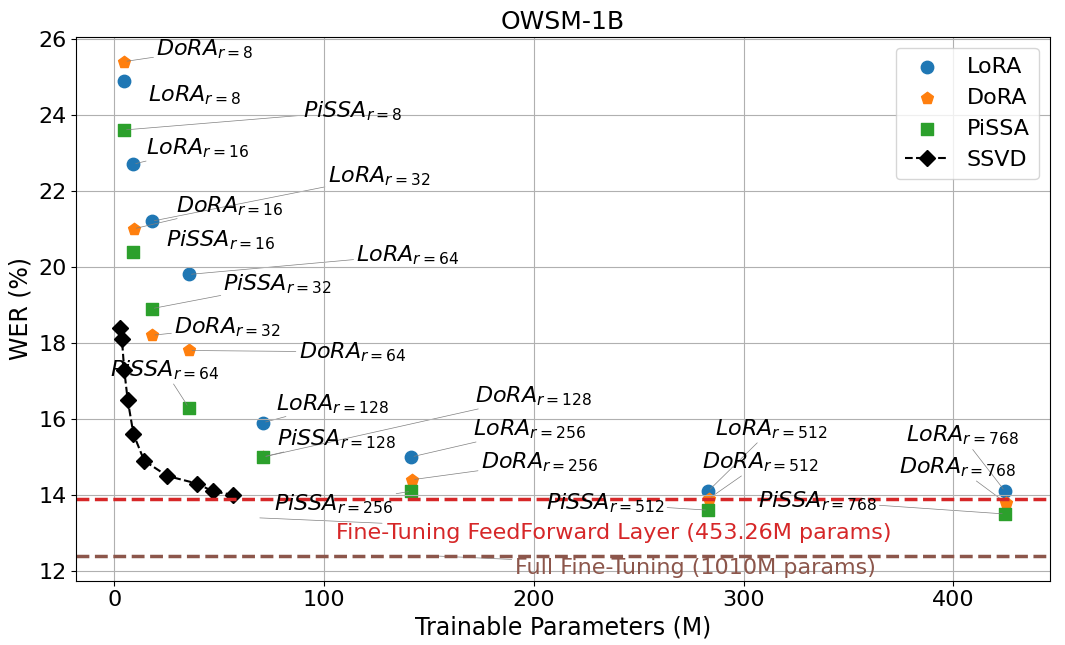}}
\caption{Word error rate (WER \%) versus trainable parameters for PEFT tuning FF layers of OWSM-1B on MyST for 10 epochs.}
\label{fig:ssvd}
\end{figure}

To address this issue, parameter-efficient fine-tuning (PEFT) methods, which aim to match full fine-tuning performance with fewer trainable parameters, have been actively studied with language modeling (LM)~\cite{houlsby2019parameter,hu2022lora}. Especially, low-rank adaptation (LoRA)~\cite{hu2022lora} has gained wide adoption due to its reduction of trainable parameters without 
introducing additional inference latency. While LoRA has been increasingly used in the speech domain, its state-of-the-art (SoTA) variants, such as weight-decomposed LoRA (DoRA)~\cite{liu2024dora} and principal singular values and singular vectors adaptation (PiSSA)~\cite{meng2024pissa}, which further improve adaptation efficiency in LM tasks, remain underexplored in speech applications. 

In~\cite{wang2025ssvd}, we demonstrated that domain-shift issues in ASR persist with the current LoRA method and its SoTA variants. Unlike text-based tasks, ASR involves a \textit{multi-modal (speech) input-(text) output structure}.
This differs fundamentally from LM, which operates within a single text-to-text modality. For example, children may pronounce "rabbit" as \textipa{[w\ae bIt]} instead of \textipa{[r\ae bIt]} due to developmental articulation, creating a divergence between the spoken form and its textual transcription. Therefore, in ASR fine-tuning, it is critical to \textit{separately} address transformations in input (speech) and output (text) spaces, while LoRA and its SoTA variants update both the input and output feature space equally, which inevitably reduces adaptation efficiency. 
In~\cite{wang2025ssvd}, we introduced a structured singular value decomposition (SVD)-guided PEFT method, named SSVD, which decomposes 
initial weights into right singular vectors (representing the input feature space) and left singular vectors (capturing the output semantic feature space). SSVD updates right singular vectors to align better with domain-shifted speech inputs, while keeping the left singular vectors \textit{fixed} to preserve semantic mappings.  
 
In~\cite{wang2025ssvd}, we benchmarked LoRA, VeRA~\cite{kopiczko2024vera}, DoRA, PiSSA, SVFT~\cite{lingam2024svft}, and SSVD on two domain-shift ASR tasks: child speech and regional accents. We showed that SSVD achieved higher efficiency than other SoTA methods with very few trainable parameters, although a performance gap to full fine-tuning remained. Therefore, in Figure~\ref{fig:ssvd}, we extend the comparison of word error rate (WER) by including the best-performing PEFT methods from~\cite{wang2025ssvd}, fine-tuned on the feedforward (FF) layers of the OWSM-1B model~\cite{peng2024owsm} using the MyST child speech dataset~\cite{pradhan2024myst}. We scale the PEFT parameter size up toward the full fine-tuning level to explore the gap between full fine-tuning and PEFT methods. This reveals three unresolved issues: (1) \textbf{Scalability of SSVD}: When only the input space is adapted (termed \textit{inner transformation}), SSVD shows lower WER with very few trainable parameters. However, it lacks scalability compared to other methods. Even when tuning 100\% of the right singular vectors, only 56.68M parameters are involved, lagging behind other PEFT methods with over 280M parameters. (2) \textbf{Understanding parameter budget efficiency}: While experimental results indicate that SSVD with inner transformation outperforms other methods at very small parameter budgets, the reasons behind this advantage remain unclear. This relates directly to a fundamental open question in PEFT: \textit{how should limited parameter or compute budgets be allocated?} For instance, is it more effective to tune parameters associated with the input space, the output space, or both? (3) \textbf{Balancing learning and forgetting under constrained budgets}: Parameter allocation affects both adaptation capacity and the extent of forgetting. Understanding how different allocation strategies impact the trade-off between learning and forgetting remains an open question.

To address these issues, we extend the SSVD method by incorporating both the input (speech)-associated inner transformation and an output (text) space–associated outer transformation, resulting in a scalable tuning framework named SSVD-O. We conduct an in-depth analysis of parameter allocation by varying the extent of inner and outer transformations across domain-shifted ASR tasks, including low-resource child speech and regional accents (Dutch as spoken in The Netherlands or Belgium (=Flemish)). Our experiments fine-tune OWSM models at scales from 0.1B to 2B using the widely adopted open-source ESPnet toolkit. SSVD-O is compared to LoRA and its SoTA variants, DoRA and PiSSA, from very small to nearly full fine-tuning
parameter budgets, 
in order to investigate the performance gap between PEFT and full fine-tuning. Additionally, we evaluate the adapted models on adult speech and multilingual corpora used during OWSM pretraining to analyze the degree of forgetting introduced by each method.

The main contributions of this paper are:
\begin{itemize}[itemsep=0pt, topsep=2pt, leftmargin=*]
\item We extend the structured SVD-guided (SSVD) PEFT method, originally limited to input-associated inner transformation, into a fully adaptable framework by introducing an output-associated outer transformation. The resulting new method, SSVD-Outer (SSVD-O), enables scalable PEFT tuning and achieves improved performance over fine-tuning and SoTA PEFT methods, including LoRA, DoRA, PiSSA, and the original SSVD, on domain-shifted ASR tasks, including child speech and regional accents.
\item We conduct the first systematic study on parameter budget allocation in PEFT for speech recognition. By benchmarking SSVD(-O) against LoRA, DoRA, and PiSSA within the ESPnet framework across model scales from 0.1B to 2B, we analyze how tuning different subspaces (input vs. output) impacts adaptation efficiency under strict parameter constraints, providing practical guidance for efficient PEFT design.
\item We provide an in-depth analysis of catastrophic forgetting in PEFT by comparing SSVD-O with full fine-tuning, LoRA, DoRA, PiSSA, and SSVD across two domain-shift scenarios: child-to-adult speech and dialect-to-multilingual speech. Our findings reveal how input- and output-associated subspaces contribute differently to preserving generalization across domains, guiding future work in continual learning.
\end{itemize}

\section{SSVD-O Method}
\label{sec:method}
In this section, we first recap the SSVD method and then introduce its extended outer transformation, SSVD-O. \underline{Trainable variables} are indicated by \underline{underlining}.

Similar to LoRA, SSVD freezes the pre-trained model weights and injects trainable low-rank (usually $r$ in LoRA) matrices into each linear layer during fine-tuning. The parameter updates are guided by the SVD of the pre-trained weight matrix. Mathematically, for a pre-trained weight $\mathbf W_0\in\mathbb{R}^{m\times n}$, with $m \geq n$ (e.g., a feedforward layer), its SVD is given by:
\vspace{-0.5em}
\begingroup
\setlength{\abovedisplayskip}{0.5em}
\setlength{\belowdisplayskip}{0.5em}
\setlength{\abovedisplayshortskip}{0pt}
\setlength{\belowdisplayshortskip}{0pt}
\begin{equation}
\label{eq:svd}
\mathbf{W}_0 = \mathbf{U} \boldsymbol{\Sigma} \mathbf{V}^{\top} = \sum_{i=1}^{n} \sigma_i \mathbf{u}_i \mathbf{v}_i^\top,
\end{equation}
\endgroup
where, $\mathbf U\in\mathbb{R}^{m\times n}$, $\mathbf V\in\mathbb{R}^{n\times n}$ contain the left and right singular vectors $\mathbf{u}_i$ and $\mathbf{v}_i$ respectively, and $\boldsymbol{\Sigma}\in\mathbb{R}^{n\times n}$ is a diagonal matrix of singular values $\sigma_i$. For each linear mapping, an input $\mathbf x\in\mathbb{R}^n$, represented in the coordinate system spanned by the right singular vector $\mathbf v_i$, is mapped to an output $\mathbf y\in\operatorname{Col}\mathbf{W}_0\subset\mathbb{R}^{m}$ in the coordinate system spanned by the left singular vector $\mathbf u_i$. 
\begingroup
\setlength{\abovedisplayskip}{0.5em}
\setlength{\belowdisplayskip}{0.5em}
\setlength{\abovedisplayshortskip}{0pt}
\setlength{\belowdisplayshortskip}{0pt}
\begin{equation}
\mathbf{y} = \mathbf{U} \boldsymbol{\Sigma} \mathbf{V}^{\top}\mathbf{x} 
\end{equation}
\endgroup

\textbf{SSVD} focuses on the shift of the input feature $\mathbf{x}$ and structurally adapts the input space by rotating and scaling the right singular vector basis $\mathbf{v}_i$ towards a shifted basis $\mathbf{v}_i'$ (the \textit{inner transform}):
\begingroup
\setlength{\abovedisplayskip}{0.5em}
\setlength{\belowdisplayskip}{0.5em}
\setlength{\abovedisplayshortskip}{0pt}
\setlength{\belowdisplayshortskip}{0pt}
\begin{equation}
\label{eq:ssvd}
\mathbf{y} = \mathbf{U} \boldsymbol({\Sigma + \underline{\mathbf{\Delta \Sigma}}})\underline{\mathbf{G}}\mathbf{V}^{\top}\mathbf{x}\\
           = \mathbf{W}'\mathbf{x}
\end{equation}
\endgroup
where diagonal matrix $\underline{\Delta\boldsymbol{\Sigma}}$ models axis-wise scaling (i.e., singular value shifts), and $\underline{\mathbf{G}} \in \mathbb{R}^{n \times n}$ is an orthogonal matrix representing a series of rotations in the right singular vector space. Leveraging the Eckart–Young theorem~\cite{eckart1936approximation}, SSVD restricts adaptation to the top-$k$ singular components for efficiency:
\begin{equation}
\label{eq:ssvd-k}
\underline{\mathbf{\Delta \Sigma}} = \begin{bmatrix} \underline{\mathbf{\Delta \Sigma}_k} & 0 \\ 0 & 0\end{bmatrix}~~\text{and}~~\underline{\mathbf{G}} = \begin{bmatrix} \underline{\mathbf{G}_k} & 0 \\ 0 & I \end{bmatrix}
\end{equation}
Therefore, the number of trainable parameters in SSVD is limited to $\frac{k(k+1)}{2}$, where $k\leq n$, which is significantly fewer than that of LoRA, DoRA or PiSSA. In comparison, the maximum parameter count of LoRA or PiSSA is $r\times(m+n)$ with low-rank $r=n$, while DoRA requires even more, resulting in over twice the size of the full-rank SSVD configuration.

\textbf{SSVD-O} further enhances model flexibility by jointly adapting the output feature space. It approximates a rotation (the \textit{outer transform}) of the left singular vector basis $\mathbf{U}$ by adding its orthonormal complement $\mathbf{U}_2 \in \mathbb{R}^{m \times (m-n)}$ obtained from full SVD of $\mathbf{W}_0$:
\begin{equation}
\label{eq:ssvd-o}
\mathbf{y}
= \bigl(\mathbf{U} + \mathbf{U}_2\begin{bmatrix} \underline{\mathbf{Q}} & \mathbf{0} \end{bmatrix} \bigr)
\bigl(\boldsymbol{\Sigma} + \underline{\Delta\boldsymbol{\Sigma}}\bigr)\underline{\mathbf{G}}\mathbf{V}^\top \mathbf{x}
= \mathbf{W}'\mathbf{x}
\end{equation}
where $\underline{\mathbf{Q}} \in \mathbb{R}^{(m-n) \times l}$ indicates the first $l$ left singular vectors to be tuned from the original basis. To maintain orthogonality of the shifted $\mathbf{U}'=\mathbf{U} + \mathbf{U}_2\begin{bmatrix} \underline{\mathbf{Q}} & \mathbf{0} \end{bmatrix}$, we regularize:
\begingroup
\setlength{\abovedisplayskip}{0.5em}
\setlength{\belowdisplayskip}{0.5em}
\setlength{\abovedisplayshortskip}{0pt}
\setlength{\belowdisplayshortskip}{0pt}
\begin{equation}
\mathbf{U}_l'^{\top}\mathbf{U}_l' =  \mathbf{I} + \underline{\mathbf{Q}}^\top\underline{\mathbf{Q}} \approx \mathbf{I},\qquad \mathbf{U}_l'= \mathbf{U}'_{[:, :l]}
\end{equation}
\endgroup
with a constraint:
\begingroup
\setlength{\abovedisplayskip}{0.5em}
\setlength{\belowdisplayskip}{0.5em}
\setlength{\abovedisplayshortskip}{0pt}
\setlength{\belowdisplayshortskip}{0pt}
\begin{equation}
\label{eq:q_constraint}
\underline{\mathbf{Q}}^\top\underline{\mathbf{Q}} \approx \tau\mathbf{I},\qquad 0<\tau\ll 1
\end{equation}
\endgroup
This ensures that $\mathbf{U}_l'^\top\mathbf{U}_l'\approx (1+\tau)\mathbf{I}$ preserving approximate orthonormality up to a small correction.
For stable and efficient training, 
the constraint~(\ref{eq:q_constraint}) is implicitly enforced by parameterizing $\underline{\mathbf{Q}}$ via Cholesky decomposition without introducing regularization loss:
\begingroup
\setlength{\abovedisplayskip}{0.5em}
\setlength{\belowdisplayskip}{0.5em}
\setlength{\abovedisplayshortskip}{0pt}
\setlength{\belowdisplayshortskip}{0pt}
\begin{equation}
\underline{\mathbf{Q}} = \underline{\mathbf{L}} \underline{\mathbf{R}}^{-\top}, \quad \text{where} \quad \underline{\mathbf{R}} = \operatorname{Cholesky}(\underline{\mathbf{L}}^\top \underline{\mathbf{L}} + \tau \mathbf{I})
\end{equation}
\endgroup
Therefore, $\underline{\mathbf{Q}}$ in~(\ref{eq:ssvd-o}) is generated by unconstrained $\underline{\mathbf{L}} \in \mathbb{R}^{(m-n) \times l}$.

\begin{figure*}[!tb]
\centerline{\includegraphics[width=0.9\linewidth]{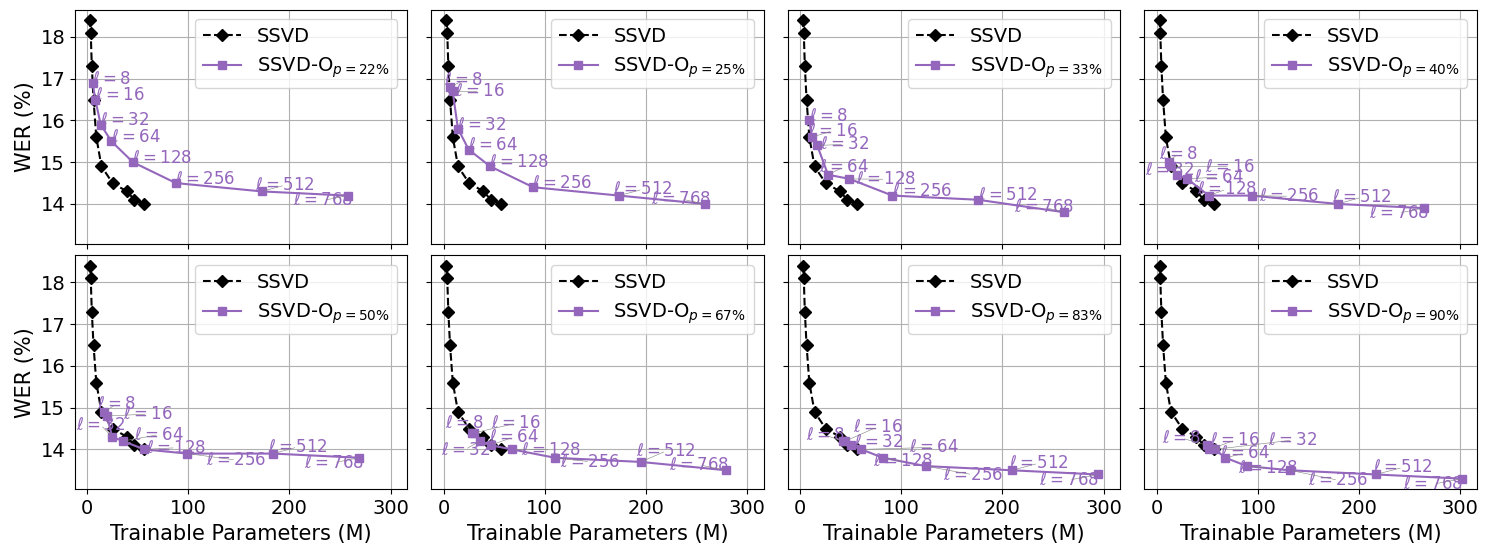}}
\caption{WER (\%) versus number of trainable parameters for SSVD and SSVD-O with varying combinations of inner transformation ratio $p\in [22\%,100\%]$ and outer transformation rank $l\in[8, 708]$, evaluated on OWSM-1B fine-tuned on the MyST dataset for 10 epochs.}
\label{fig:ssvd_budget_1B}
\end{figure*}
\begin{figure*}[!tb]
\centerline{\includegraphics[width=0.9\linewidth]{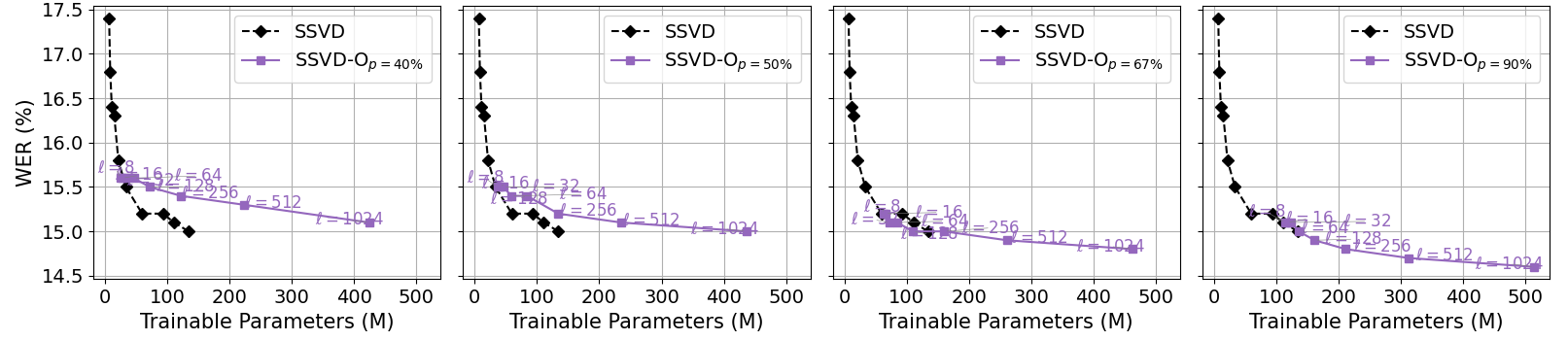}}
\caption{WER (\%) versus number of trainable parameters for SSVD and SSVD-O with varying combinations of inner transformation ratio $p\in [22\%,100\%]$ and outer transformation rank $l\in[8, 1024]$, evaluated on OWLS-2B fine-tuned on the MyST dataset for 5 epochs.}
\label{fig:ssvd_budget_2B}
\end{figure*}
\begin{figure}[!tb]
\centerline{\includegraphics[width=1.0\linewidth]{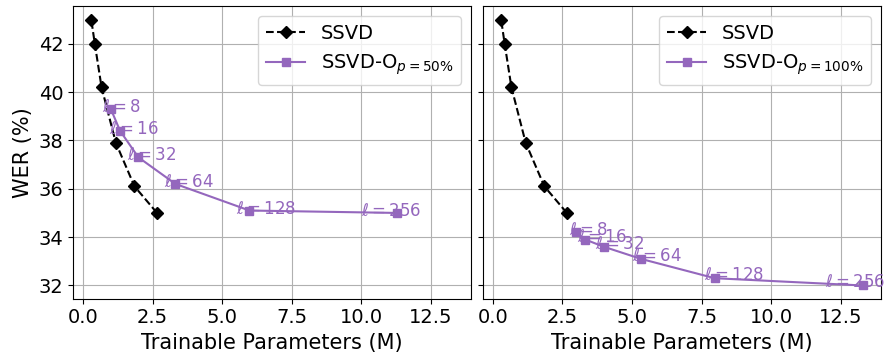}}
\caption{WER (\%) versus number of trainable parameters for SSVD and SSVD-O with varying combinations of inner transformation ratio $p\in [33\%,100\%]$ and outer transformation rank $l\in[8, 256]$, evaluated on OWSM-0.1B fine-tuned on the CGN dataset for 5 epochs.}
\label{fig:ssvd_budget_0.1B}
\end{figure}
\captionsetup{font=small}
\begin{figure}[!tb]
\centerline{\includegraphics[width=0.9\linewidth]{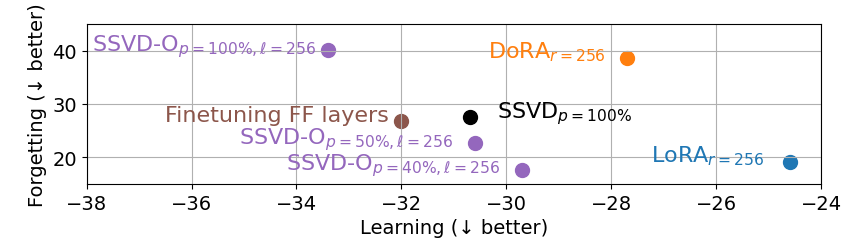}}
\caption{Learning vs. Avg. forgetting from Table~\ref{tab:forgetting_0.1B}, \textit{the bottom-left indicates the best trade-off}.}
\label{fig:forgetting_0.1B}
\end{figure}
\section{Experiments}
\label{sec:experiment}
We adopt the open-source Whisper-style speech models (OWSM and OWLS)~\cite{peng2023reproducing, peng2024owsm, chen2025owls} as base models due to their transparent training corpora~\cite{chen2025owls, peng2024owsm}, which avoid domain overlap with our evaluation data. OWSM employs an E-Branchformer~\cite{kim2023branchformer} encoder and a Transformer~\cite{vaswani2017attention} decoder, while OWLS uses a standard Transformer encoder–decoder architecture. All compared PEFT methods, \textbf{LoRA}, \textbf{DoRA}, \textbf{PiSSA}, \textbf{SSVD} and \textbf{SSVD-O}, are implemented within the ESPnet framework. We evaluate them by fine-tuning all FF layers of OWSM-0.1B, OWSM-1B, and OWLS-2B on 
MyST~\cite{pradhan2024myst} (child speech) and CGN~\cite{oostdijk2000cgn} (Flemish and Dutch).

\textbf{MyST} corpus~\cite{pradhan2024myst} contains English dialogues between elementary students and virtual science tutors, with verbatim transcriptions including disfluencies. Following~\cite{attia2024kidwhisper}, we discard utterances with WER above 50\% (Whisper-large-v2), yielding 179 hours of speech. \textbf{CGN}~\cite{oostdijk2000cgn} corpus contains diverse speaking styles in Dutch and Flemish, including read speech, interviews, and spontaneous dialogues. We use the 341-hour subset 
with a 2:1 Dutch–Flemish ratio, excluding the spontaneous components $\mathtt{c}$, $\mathtt{d}$, and $\mathtt{f}$. 

To assess forgetting, we evaluate MyST-adapted models on the adult LibriSpeech test sets, and CGN-adapted models on Multilingual LibriSpeech (MLS)~\cite{pratap2020mls}, covering German (DE), Spanish (ES), French (FR), Italian (IT), Dutch (NL), Polish (PL), and Portuguese (PT). 
Both corpora were part of OWSM pretraining~\cite{peng2023reproducing, peng2024owsm, chen2025owls}.
\begin{figure*}[!tb]
\centerline{\includegraphics[width=0.93\linewidth]{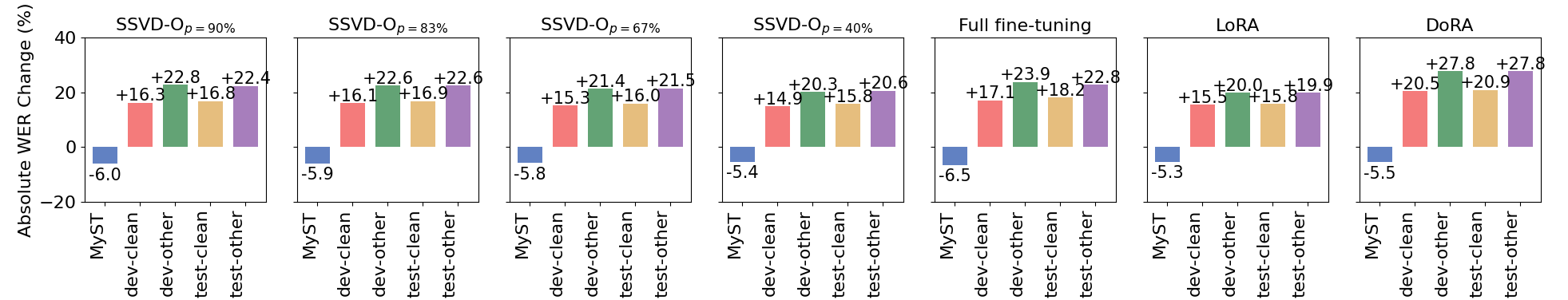}}
\caption{Absolute WER change on LibriSpeech for PEFT tuning OWSM-1B on MyST. \textit{`$+$' indicates forgetting, `$-$' indicates learning}.}
\label{fig:forgetting_1B}
\end{figure*}

For PEFT methods, we denote low-rank configurations as LoRA$_{r=\text{rank}}$, DoRA$_{r=\text{rank}}$, and PiSSA$_{r=\text{rank}}$. For SSVD and SSVD-O, we vary the number of adapted singular components: $k$ in Eq.~(\ref{eq:ssvd-k}) controls the proportion $p\%$ of adapted right singular vectors/values, denoted as SSVD$_{p=\text{portion}}$; $l$ in Eq.~(\ref{eq:ssvd-o}) specifies the number of adapted left singular vectors in SSVD-O, denoted as SSVD-O$_{p=\text{portion}, l=\text{rank}}$. Experiments use a single NVIDIA V100 32GB, A100 80GB, or H100 80GB depending on model size.
\section{Results}
\label{sec:results}
\begin{figure}[!tb]
\centerline{\includegraphics[width=0.89\linewidth]{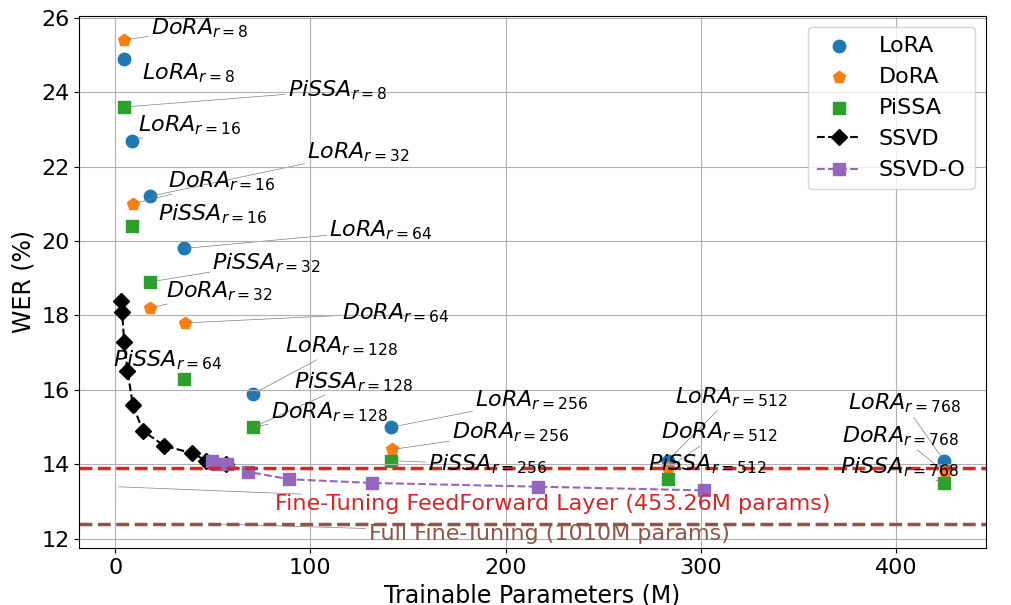}}
\caption{WER (\%) versus trainable parameters for PEFT tuning FF layers of OWSM-1B on MyST for 10 epochs.}
\label{fig:ssvd_full_1B}
\end{figure}
\begin{figure}[!tb]
\centerline{\includegraphics[width=0.9\linewidth]{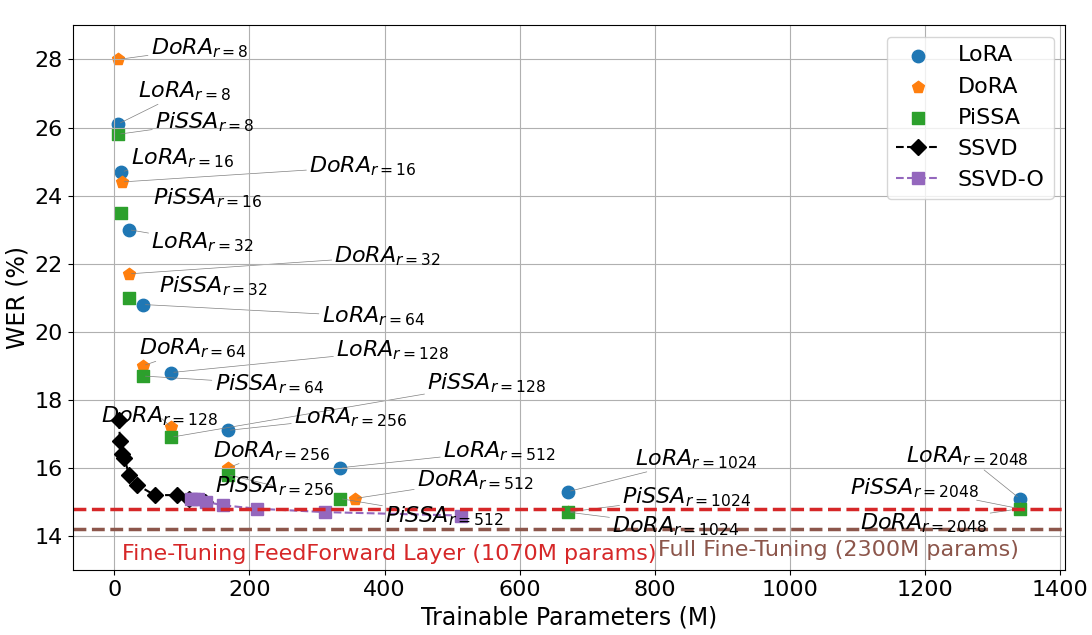}}
\caption{WER (\%) versus trainable parameters for PEFT tuning FF layers of OWLS-2B on MyST for 5 epochs.}
\label{fig:ssvd_full_2B}
\end{figure}
\vspace{-0.5em}
\begin{table}[!htbp]
\centering
\small
\caption{Absolute WER change on MLS for PEFT tuning OWSM-0.1B on CGN. \textit{`$+$' indicates forgetting, while `$-$' indicates learning}.}
\label{tab:forgetting_0.1B}
\resizebox{0.49\textwidth}{!}{
\begin{tabular}{lcc|ccccccc}
\toprule
\textbf{Model} & \textbf{CGN} & \textbf{NL} & \textbf{DE} & \textbf{ES} & \textbf{FR}& \textbf{IT} & \textbf{PL} & \textbf{PT} & \textbf{Avg. forgetting}\\
\midrule
Full fine-tuning & $\mathbf{-38.6}$ & $\mathbf{-8.4}$ & $+85.3$ & $+71.5$ & $+70.1$ & $+66.2$ & $+100.9$ & $+74.7$ & $+78.1$\\
Fine-tuning FF layers & $-32.0$ & $-5.6$ & $+37.3$ & $+10.3$ & $+12.7$ & $+19.9$ & $+51.8$ & $+28.5$ & $+26.8$ \\
\midrule
SSVD-O$_{p=100\%, l=256}$ & $\mathbf{-33.4}$ & $\mathbf{-6.1}$ & $+56.3$ & $+20.6$ & $+19.3$ & $+31.7$ & $+67.9$ & $+44.8$ & $+40.1$\\
SSVD-O$_{p=50\%, l=256}$ & $-30.6$ & $-5.1$ & $+30.0$ & $+8.3$ & $+10.0$ & $+16.8$ & $+42.4$ & $+28.0$ & $+22.6$ \\
SSVD-O$_{p=40\%, l=256}$ & $-29.7$ & $-5.0$ & $+24.7$ & $\mathbf{+6.7}$ & $\mathbf{+7.9}$ & $\mathbf{+12.6}$ & $\mathbf{+31.5}$ & $\mathbf{+22.9}$ & $\mathbf{+17.7}$\\
SSVD$_{p=100\%}$ & $-30.7$ & $-4.5$ & $+38.0$ & $+10.1$ & $+12.7$ & $+18.8$ & $+50.3$ & $+35.0$ & $+27.5$\\
\midrule
DoRA$_{r=256}$ &  $-27.7$ & $-3.0$ & $+42.8$ & $+19.9$ & $+20.3$ & $+33.7$ & $+71.0$ & $+44.5$ & $+38.7$\\
LoRA$_{r=256}$ &  $-24.6$ & $-2.9$ & $\mathbf{+18.4}$ & $+6.8$ & $\mathbf{+7.9}$ & $+15.3$ & $+38.6$ & $+28.0$ & $+19.2$\\
\bottomrule
\end{tabular}
}
\end{table}
\vspace{-0.7em}
We first discuss the \textbf{parameter budget efficiency} by comparing the WER results of SSVD and SSVD-O configured with different combinations of inner transformation ratios and outer transformation ranks. Results for fine-tuning on OWSM-1B and OWLS-2B with MyST data are shown in Figure~\ref{fig:ssvd_budget_1B} and~\ref{fig:ssvd_budget_2B}, while results on OWSM-0.1B with CGN data are presented in Figure~\ref{fig:ssvd_budget_0.1B}. The three plots show a consistent conclusion: inner transformation is more beneficial than outer transformation, especially when the parameter budget is tight. Given the same number of trainable parameters, increasing the inner transformation portion improves performance more effectively than increasing the outer transformation rank. This explains why, in Figure~\ref{fig:ssvd}, SSVD outperforms other methods with significantly fewer trainable parameters, as the others rely on both inner and outer transformations. The effectiveness of outer transformation becomes more pronounced as the model size increases, contributing additional gains in larger-scale settings. 

With the extended outer transformation, we \textbf{scale SSVD} and report WER for tuning FF layers of OWSM-1B and OWLS-2B on MyST in Figure~\ref{fig:ssvd_full_1B} and~\ref{fig:ssvd_full_2B}. In both model scales, SSVD-O consistently outperforms fine-tuning all FF layers while using fewer parameters. Its performance typically lies between FF-layer fine-tuning and full fine-tuning, further validating the effectiveness of the outer transformation, especially in larger models. Note that OWLS-2B is shallower and wider than OWSM-1B, its worse zero-shot performance and fewer training epochs may explain the overall performance gap.
To evaluate \textbf{the trade-off between learning and forgetting}, we report absolute WER changes on four LibriSpeech test sets after PEFT tuning FF layers of OWSM-1B model on MyST(Figure~\ref{fig:forgetting_1B}). Positive values indicate degradation due to forgetting, while negative values reflect error reduction by learning. In both cases, lower is better. 
Based on experimental observations (not included here), SSVD-O with higher outer transformation ranks reduces WER and forgetting compared to SSVD without outer adaptation. Thus, SSVD-O results in Figure~\ref{fig:forgetting_1B} use outer rank $l=768$. 
We observe that smaller inner transformation ratios tend to yield less forgetting, this is also supported by Table~\ref{tab:forgetting_0.1B}, which reports WER changes on MLS after tuning OWSM-0.1B on CGN.
Therefore, using a smaller inner transformation ratio combined with a larger outer transformation rank achieves a better trade-off between learning and forgetting, such as SSVD-O${_p=40\%}$ (Figure~\ref{fig:forgetting_1B}) and SSVD-O$_{p=50\%, l=256}$ (Table~\ref{tab:forgetting_0.1B}).
This property could be exploited further with methods in \textit{continual learning tasks}~\cite{wang2024comprehensive}.
Table~\ref{tab:forgetting_0.1B} also summarizes learning vs. forgetting across languages, with the best learning and least forgetting shown in \textbf{bold}. Overall, 
SSVD-O achieves less forgetting than other methods with similar or even higher learning ability. The trade-off is visualized in Figure~\ref{fig:forgetting_0.1B}. Optimal performance is found in the bottom-left.

\section{Conclusion}
In this work, we propose SSVD-O, a structured SVD-guided PEFT method that leverages the multi-modal nature of speech-to-text by separately adapting input speech-associated inner transformations and output semantic-associated outer transformations. We systematically analyze parameter budget efficiency and show that SSVD-O outperforms fine-tuning and SoTA LoRA variants on domain-shifted ASR tasks, including child speech and regional accents. Our learning–forgetting analysis further shows that SSVD-O achieves the best trade-off between high learning ability and less forgetting.

\vfill\pagebreak

\section*{Acknowledgment}
Experiments of this work used the Bridges2 system at PSC and Delta system at NCSA through allocations CIS210014 and IRI120008P from the Advanced Cyberinfrastructure Coordination Ecosystem: Services \& Support (ACCESS) program, supported by National Science Foundation grants \#2138259,\#:2138286, \#:2138307, \#:2137603, and \#:2138296. \\
This research was supported by the Flemish Government under ``Onderzoeksprogramma AI Vlaanderen'', the FWO-SBO grant S004923N: NELF, the FWO grant V401325N and the computational resources and services used in this work were provided by the VSC (Flemish Supercomputer Center), funded by the Research Foundation Flanders (FWO) and the Flemish Government – department WEWIS (2025-85).
\printbibliography
\end{document}